\documentclass[12pt]{article}
\usepackage{graphicx}
\usepackage{epstopdf}
\usepackage{amssymb,amsmath,epsfig}

\begin{document}

\title{\bf On Cracking of Charged Anisotropic Polytropes}

\author{M. Azam$^{1}$ \thanks{azam.math@ue.edu.pk, azammath@gmail.com} and
S. A. Mardan$^2$ \thanks{syedalimardanazmi@yahoo.com, ali.azmi@umt.edu.pk}\\
$^1$ Division of Science and Technology, University of Education,\\
Township Campus, Lahore-54590, Pakistan.\\
$^2$ Department of Mathematics,\\ University of the Management and Technology,\\
C-II, Johar Town, Lahore-54590, Pakistan.}

\date{}

\maketitle
\begin{abstract}
Recently in \cite{34}, the role of electromagnetic field
on the cracking of spherical polytropes has been investigated
without perturbing charge parameter explicitly.
In this study, we have examined the occurrence of cracking
of anisotropic spherical polytropes
through perturbing parameters like anisotropic pressure, energy density and charge.
We consider two different types of polytropes in this study.
We discuss the occurrence of cracking in two different ways
$(i)$ by perturbing polytropic constant, anisotropy and charge parameter
$(ii)$ by perturbing polytropic index, anisotropy and
charge parameter for each case. We conclude that cracking appears
for a wide range of parameters in both cases. Also, our results are
reduced to \cite{33} in the absence of charge.
\end{abstract}

\section{Introduction}

Polytropes plays very important role for the description and modeling of stellar
structure of compact objects. Many scientists have been involved in the
study of polytropes due to lucid equation of state (EoS) and resulting
Lane-Emden equations which helped us understanding various
compact objects phenomenon. In Newtonian regime, Chandrasekhar \cite{1}
was the founder of theory of polytropes emerging through laws of
thermodynamics. Tooper [2,3] was the first who developed the basic
frame work for polytropes under the assumption of quasi-static equilibrium
condition for the derivation of Lane-Emden equations. Kovetz \cite{4}
identifies some errors in the work of Chandrasekhar \cite{1} and provided new
corrected results which revolutionized the theory of slowly rotating
polytropes. Abramowicz \cite{5} initially presented the concept of higher dimensional
polytropes and presented the modified form of Lane–Emden equations.

The discussion of charge in the modeling of relativistic objects always
attracts the researchers. Bekenstein \cite{6} formulated the the
hydrostatic equilibrium equation which helped in the study of gravitational collapse of stars.
Bonnor \cite{7,8} showed that electric repulsion might affect
gravitation collapse in stars. The contraction of charged compact
stars in isotropic coordinates was studied by Bondi \cite{9}.
Koppar et al. \cite{10} presented a novel scheme to find out charge generalization
in compact relativistic stars with static inner charged
fluid distribution. Ray et al. \cite{11} discussed higher
density stars and concluded that they can hold very large
amount of charge which is approximately $10^{20}$ coulomb.
Herrera et al. \cite{12} studied charged spherical compact
objects with dissipative inner fluid distribution by means
of structure scalars. Takisa \cite{13} developed the models
of charged polytropic stars.

The role of anisotropy is vital in the discussion of stellar structure
of stars and many physical phenomena cannot be described without it.
Cosenza et al. \cite{14} developed a heuristic way for the modeling
of stars with anisotropic fluid distribution. Santos \cite{15}
developed the general models for Newtonian and relativistic
self-gravitating compact objects. Herrera and Barreto \cite{16}
described relativistic polytropes and developed a new methodology
of effective variables to elaborate physical variables involved
in polytropic models. Herrera et al. \cite{17} formulated governing
equations with anisotropic stress for self-gravitating spherically
symmetric distributions. Herrera and Barreto \cite{18,19} described
a new way to study stability of polytropic models by means of
Tolman-mass (which is the measure of active gravitational mass).
Herrera et al. \cite{20} used conformally flat condition which
reduced the polytropic model parameters and hence provide better
understanding of spherical polytropes in relativistic regime.

In the process of mathematical modeling of stellar structure,
the stability analysis of developed model is essential for its
validity and future implementation. A model is worthless if it
is unstable towards small fluctuations in model parameters.
In this regard, Bondi \cite{21} initially
proposed hydrostatic equilibrium equation for the analysis of stability of
spherical models. Herrera \cite{22} presented the concept of
cracking (overturning) to elaborate the behavior of stellar
fluid just after loosing its equilibrium state through
global density perturbation. Gonzalez et al. \cite{23,24}
extended his work by developing local density perturbation
technique. Azam et al. [25-29] applied the local density
perturbation scheme to check the stability of various
compact object models. Sharif and Sadiq \cite{30} developed general
framework for charged polytropes and discussed their stability
by numerical approach. Azam et al. \cite{31} presented the
theory of charged spherical polytropes with generalized
polytropic EoS and provide the stability analysis by means
of bounded Tolman mass function. They also extended this idea
for cylindrical polytropes and checked the stability through
Whittaker mass \cite{32}. Herrera et al. \cite{33} analyzed
the stability of spherical polytropes by means of cracking
(or overturning) by applying perturbation in local anisotropy
and energy density. Sharif \cite{34} applied the same technique
presented by Herrera \cite{33} for charged spherical polytropes.

The plan of this work is as follows. In section \textbf{2},
we describe some basic equations and conventions.
Section \textbf{3} and \textbf{4} are devoted for the discussion of polytropes
of case \textbf{1} and \textbf{2}, respectively. In the last section, we conclude our results.

\section{Basic Equations and Conventions}
{
\allowdisplaybreaks
We consider static spherically symmetric space time
\begin{equation}\label{1}
ds^2=-e^{\nu}dt^{2}+e^{\lambda}dr^{2}+r^2 d\theta^{2}+r^2 \sin^2\theta{d\phi^2},
\end{equation}
where $\nu(r)$ and $\lambda(r)$ both depends only on radial
coordinate $r$. The generalized form of energy-momentum tensor
for charged anisotropic inner fluid distribution is given by
\begin{equation}\label{2}
T_{i j}=(P_t+\rho) V_{i} V_{j} +g_{i j}P_t +(P_r - P_t) S_{i}
S_{j}+\frac{1}{4\pi}(F_{i}^{m} F_{j m}-\frac{1}{4} F^{mn} F_{mn} g_{ij}),
\end{equation}
where $P_t$, $P_r$, $\rho$, $V_i$, $S_i$ and $F_{mn}$ represent the
tangential pressure, radial pressure, energy density, four velocity,
four vector and Maxwell field tensor for the inner fluid distribution
and they satisfies following conditions
\begin{eqnarray}\label{3}\notag
&&V^{i}=e^{\frac{-\nu}{2}} \delta^i_0,~S^{i}= e^{\frac{-\lambda}{2}}
\delta_1^i,~V^{i}V_{i}=-1,~S^{i}S_{i} =1,~S^{i}V_{i} =0,\\&&
F_{ij}=\phi_{j,i}-\phi_{i,j},~F^{ij}_{;j}=\mu_0 J^i,~F_{[ij;k]}=0,
\end{eqnarray}
where $\phi_i$ is the four-potential, $\mu_0$ is the magnetic
permeability and $J^i$ is the four-current.
Moreover, $\phi_i$ and $V_i$ satisfies the following relations
\begin{equation}\label{7}
\phi_{i}=\phi(r) \delta^0_i,~~~J_i=\sigma V_i,
\end{equation}
where $\phi$ and $\sigma$ are the scalar potential and the charge
density respectively. The Maxwell field equations corresponding to (\ref{1})
yields
\begin{equation}\label{8}
\phi^{\prime\prime}+\Big(\frac{2}{r}-\frac{\nu^\prime}{2}
-\frac{\lambda^\prime}{2}\Big)\phi^\prime=4 \pi
 \sigma e^{\frac{\nu+\lambda}{2}},
\end{equation}
where $``\prime"$ denotes the differentiation with respect to $r$.
From above, we have
\begin{equation}\label{9}
\phi^{\prime}=\frac{q(r)}{r^2}e^{\frac{\nu+\lambda}{2}},
\end{equation}
where $q(r)=4 \pi \int_0^r \mu e^{\frac{\lambda}{2}} r^2 dr $
represents the total charge inside the sphere.
The Einstein-Maxwell field equations for line element (\ref{1})
are given by
\begin{eqnarray}\label{10}
\frac{\lambda^\prime
e^{-\lambda}}{r}+\frac{(1-e^{-\lambda})}{r^2}=8\pi \rho
+\frac{q^2}{r^4}, \\\label{11}
\frac{\nu^\prime e^{-\lambda}}{r}-\frac{(1-e^{-\lambda})}{r^2}=8\pi P_r
-\frac{q^2}{r^4}, \\\label{12}
e^{-\lambda} \bigg[\frac{\nu^{\prime\prime}}{2}-\frac{\nu^\prime
\lambda^\prime}{4}+\frac{\nu^{\prime^2}}{4}+\frac{\lambda^\prime
-\nu^\prime}{2r}\bigg] = 8 \pi P_t+\frac{q^2}{r^4}.
\end{eqnarray}
Solving Eqs.(\ref{10})$-$(\ref{12}) simultaneously lead to
hydrostatic equilibrium equation
\begin{equation}\label{13}
\frac{d P_r}{dr}-\frac{2}{r}\Big(\Delta+\frac{q q^\prime}
{8\pi r^3}\Big)+\Big(\frac{4\pi r^4 P_r-q^2+m r}{r(r^2-2mr+q^2)}\Big)(\rho+P_r)=0,
\end{equation}
where we have used $\Delta=(P_t-P_r)$.
We take the Reissner-Nordstr\"{a}m space-time as the exterior geometry
\begin{equation}\label{14}
ds^2=-\Big(1-\frac{2M}{r}+\frac{Q^2}{r^2}\Big)dt^2+
\Big(1-\frac{2M}{r}+\frac{Q^2}{r^2}\Big)^{-1}dr^2+r^2
d\theta^{2}+r^2 \sin^2\theta{d\phi^2}.
\end{equation}
The junction conditions are very important in mathematical modeling of
compact stars. They provide us the criterion for the
interaction of two metrics, which can results a physically viable
solution \cite{35,36}. For smooth matching of two space times, we must have
\begin{equation}\label{15}
e^{\nu}=e^{-\lambda}=\Big(1-\frac{2M}{r}+
\frac{Q^2}{r^2}\Big),~~m(r)=M,~q(r)=Q,~P_r=0,
\end{equation}
and Misner-Sharp mass \cite{37} leads to
\begin{equation}\label{16}
m(r)=\frac{r}{2}(1-e^{-\lambda}+\frac{q^2}{r^2}).
\end{equation}
All the models have to satisfy following physical requirement such as
\begin{equation}\label{17}
\rho+\frac{q^2}{8\pi r^4}>0,~~\frac{P_r}{\rho}\leqslant 1
+\frac{q^2}{4\pi\rho r^4},~~\frac{P_t}{\rho}\leqslant 1.
\end{equation}
Let us now develop an anisotropic polytropes
satisfying the hydrostatic equilibrium equation
$(\ref{13})$. Furthermore, fluid distribution satisfies
the following equation
\begin{equation}\label{18}
\Delta=C(\rho+P_r)\Big[\frac{4\pi r^4 P_r-q^2+m r}
{r^2-2mr+q^2}\Big],
\end{equation}
where $C$ is a constant, produce hydrostatic equilibrium equation
\begin{equation}\label{19}
R\equiv\frac{d P_r}{dr}+h(\rho+P_r)\Big(\frac{4\pi r^4 P_r-q^2+m r}
{r(r^2-2mr+q^2)}\Big)-\frac{q}{4\pi r^4}\frac{d q}{dr}=0,
\end{equation}
with $h=1-2C$.

\section{Charged Anisotropic Polytropes}

In this section, we will analyze the
cracking of charged relativistic anisotropic polytropes through
perturbation on parameters involve in the model via two different cases.

\subsection{Polytropes for case 1}

In this section, we evaluate Eq. (\ref{19}) to obtained Lane-Emden equation.
For this, we consider the ploytropic EoS as
\begin{equation}\label{20}
P_r=K\rho_o^{1+\frac{1}{n}},
\end{equation}
so that the original polytropic part remain conserved. Also,
the mass density $\rho_o$ is related to total energy density
$\rho$ as \cite{7}
\begin{equation}\label{21}
\rho=\rho_{o}+n P_r.
\end{equation}
Now taking following assumptions
\begin{equation}\label{22}
\alpha=\frac{P_{rc}}{\rho_{c}},~~~
r=\frac{\xi}{A},~~~
\rho_{o}=\rho_{oc}\psi_0^n,~~~
m(r)=\frac{4\pi\rho_{c} v({\xi})}{A^3},~~~
A^2=\frac{4\pi\rho_{c}}{(n+1)\alpha},
\end{equation}
where subscript $c$ represents the values at the center of the star and
$\xi$, $\theta$ and $v$ are dimensionless variables.
Using above assumptions along with EoS $(\ref{20})$,
the new hydrostatic equilibrium equation (\ref{19}) implies
\begin{eqnarray}\label{23} \notag
&&\frac{d \psi_0}{d \xi}+\frac{h}{\xi^3}(1-n\alpha+(n+1)\alpha\psi_o)
\Big[\frac{v(\xi)\xi+\alpha \xi^4 \psi_o^{n+1}-\frac{4\pi \rho_c q^2}
{\alpha^2 (n+1)^2}}{1-2\alpha (n+1)\frac{v(\xi)}{\xi}+\frac{4\pi\rho_c q^2}
{\alpha (n+1)\xi^2}}\Big]\\&&-\frac{4\pi\rho_c q }{\alpha^3 (n+1)^3\xi^4 }
\frac{1}{\psi_o^n}\frac{d q}{d \xi}=0.
\end{eqnarray}
Now differentiating Eq.$(\ref{16})$ with respect to $``r"$
and using the assumptions given in Eq.$(\ref{22})$, we get
\begin{eqnarray}\label{24}
\frac{dv(\xi)}{d\xi}=\xi^2\psi_o^n(1-n\alpha+n\alpha\psi_0)
+\frac{4\pi\rho_c}{\alpha^2 (n+1)^2}\Big(\frac{q}{\xi}
\frac{d q}{d \xi}-\frac{q^2}{\xi^2}\Big),
\end{eqnarray}
Now perturbing the energy density and local anisotropy via $K$, $q$ and $h$
\begin{eqnarray}\label{25}
&&K\longrightarrow \tilde{K}+\delta K,~~~~~q\longrightarrow \tilde{q}+\delta q,\\
&&h\longrightarrow \tilde{h}+\delta h,\label{26}
\end{eqnarray}
it yields
\begin{eqnarray}\label{27}
&&\tilde{P_r}=\tilde{K}\rho_o^{1+\frac{1}{n}},\\
&& \tilde{\rho_o}=\rho_o+n\beta P_r, \label{28}
\end{eqnarray}
with $\beta=\frac{\tilde{K}}{K}$ and tilde denotes the perturbed quantity.
Introducing dimensionless variable
\begin{eqnarray}\label{29}
\hat{\tilde{R}}=\frac{A}{4\pi\rho_c}R,
\end{eqnarray}
where
\begin{eqnarray}\label{30}\notag
\hat{\tilde{R}}&=&\beta \psi_o \frac{d \psi_o}{d \xi}+
\frac{\tilde{h}\psi_o}{\xi^3}(1-n\alpha+(n+1)\alpha\beta\psi_o)\\
&\times&\Big[\frac{\xi\tilde{v}-\frac{4\pi\rho_c \tilde{q}^2}
{(n+1)^2 \alpha^2}+\beta \alpha \xi^4 \psi_o^{(n+1)}}
{1-2(n+1)\alpha \frac{\tilde{v}}{\xi}+\frac{4\pi\rho_c
\tilde{q}^2}{(n+1) \alpha\xi^2}}\Big]-\frac{4\pi\rho_c }
{(n+1)^3 \alpha^3}\frac{\tilde{q}}{\xi^4}\frac{d \tilde{q}}{d \xi}.
\end{eqnarray}
The equilibrium configuration of the system allows us to write
the above equation through Taylor's expansion up to first order as
\begin{eqnarray}\label{31}\notag
\delta\hat{R}&=&\hat{\tilde{R}}(\xi, 1+\delta\beta, h+\delta h,
v+\delta v, q+\delta q)\\\notag
&=&\hat{R}(\xi, 1, h, v, q)+ \Big(\frac{\partial\hat{\tilde{R}}}
{\partial \beta}\Big) \mathrel{\mathop{\Big
|_{\beta=1,~\tilde{v}=v}}_{\mathrm{\tilde{h}=h,~\tilde{q}=q}}} \delta\beta\\\notag
&&+ \Big(\frac{\partial\hat{\tilde{R}}}{\partial \tilde{v}}\Big)
\mathrel{\mathop{\Big|_{\beta=1,~\tilde{v}=v}}_{\mathrm{\tilde{h}=h,~\tilde{q}=q}}}
 \delta v
+ \Big(\frac{\partial\hat{\tilde{R}}}{\partial \tilde{h}}\Big)
\mathrel{\mathop{\Big|_{\beta=1,~\tilde{v}=v}}_{\mathrm{\tilde{h}=h,
~\tilde{q}=q}}} \delta h\\
&&+ \Big(\frac{\partial\hat{\tilde{R}}}{\partial \tilde{q}}\Big)
 \mathrel{\mathop{\Big|_{\beta=1,~\tilde{v}=v}}_{\mathrm{\tilde{h}=
 h,~\tilde{q}=q}}} \delta q,
\end{eqnarray}
and in equilibrium state $\hat{R}(\xi, 1, h, v, q)=0$, yields
\begin{eqnarray}\label{32}\notag
\delta\hat{R}&=&\Big(\frac{\partial\hat{\tilde{R}}}{\partial \beta}\Big)
 \mathrel{\mathop{\Big|_{\beta=1,~\tilde{v}=v}}_{\mathrm{\tilde{h}=h,~\tilde{q}=q}}}
 \delta\beta
+ \Big(\frac{\partial\hat{\tilde{R}}}{\partial \tilde{v}}\Big)
 \mathrel{\mathop{\Big|_{\beta=1,~\tilde{v}=v}}_{\mathrm{\tilde{h}=h,~\tilde{q}=q}}} \delta v\\
&&+ \Big(\frac{\partial\hat{\tilde{R}}}{\partial \tilde{h}}\Big)
 \mathrel{\mathop{\Big|_{\beta=1,~\tilde{v}=v}}_{\mathrm{\tilde{h}=h,~\tilde{q}=q}}}
  \delta h+ \Big(\frac{\partial\hat{\tilde{R}}}{\partial \tilde{q}}\Big)
  \mathrel{\mathop{\Big|_{\beta=1,~\tilde{v}=v}}_{\mathrm{\tilde{h}=h,~\tilde{q}=q}}} \delta q.
\end{eqnarray}
Using $(\ref{30})$, we have calculated the perturbed quantities of above equation
\begin{eqnarray}\label{33}\notag
\Big(\frac{\partial\hat{\tilde{R}}}{\partial \beta}\Big)
\mathrel{\mathop{\Big|_{\beta=1,~\tilde{v}=v}}_{\mathrm{\tilde{h}=h,~\tilde{q}=q}}}
&=&\psi_o^n \frac{d \psi_o}{d \xi}+
\frac{h \alpha  \psi_o ^{n+2} (1-n \alpha +(1+n) \alpha
 \psi_o )}{\xi ^3 \left(1-\frac{2 (1+n) v \alpha }{\xi }+\frac{4 \pi  q^2 \rho_c }{(1+n)
\alpha  \xi ^2}\right)}\\
&+&\frac{h (1+n) \alpha  \psi_o ^2 \left(v \xi -\frac{4 \pi
  q^2 \rho_c }{(1+n)^2 \alpha ^2}+\alpha  \psi_o ^{1+n}\right)}{\xi ^3
\left(1-\frac{2 (1+n) v \alpha }{\xi }+\frac{4 \pi  q^2
\rho_c }{(1+n) \alpha  \xi ^2}\right)},
\end{eqnarray}
\begin{eqnarray}\label{34}\notag
&&\Big(\frac{\partial\hat{\tilde{R}}}{\partial \tilde{v}}\Big)
 \mathrel{\mathop{\Big|_{\beta=1,~\tilde{v}=v}}_{\mathrm{\tilde{h}=
 h,~\tilde{q}=q}}}=
\frac{h \psi_o  (1-n \alpha +(1+n) \alpha  \psi_o )}
{\xi ^2 \left(1-\frac{2 (1+n) v \alpha }{\xi }+\frac{4 \pi
q^2 \rho_c }{(1+n) \alpha \xi ^2}\right)}+\\
&&\frac{2 h (1+n) \alpha  \psi_o  (1-n \alpha +(1+n)
\alpha  \psi_o ) \left(v \xi -\frac{4 \pi  q^2
\rho_c }{(1+n)^2 \alpha ^2}+\alpha
\psi_o ^{1+n}\right)}{\xi ^4 \left(1-\frac{2 (1+n) v
 \alpha }{\xi }+\frac{4 \pi  q^2 \rho_c }{(1+n) \alpha  \xi ^2}\right)^2},
\end{eqnarray}
\begin{eqnarray}\label{35}\notag
&&\Big(\frac{\partial\hat{\tilde{R}}}{\partial \tilde{h}}\Big)
 \mathrel{\mathop{\Big|_{\beta=1,~\tilde{v}=v}}_{\mathrm{\tilde{h}=h,~\tilde{q}=q}}}
=\frac{\psi_o  (1-n \alpha +(1+n) \alpha  \psi_o )
 \left(v \xi -\frac{4 \pi  q^2 \rho_c }{(1+n)^2 \alpha ^2}+\alpha
  \psi_o ^{1+n}\right)}{\xi
^3 \left(1-\frac{2 (1+n) v \alpha }{\xi }+
\frac{4 \pi  q^2 \rho_c }{(1+n) \alpha  \xi ^2}\right)},
\end{eqnarray}
\begin{eqnarray}\label{35}\notag
&&\Big(\frac{\partial\hat{\tilde{R}}}{\partial \tilde{q}}\Big)
 \mathrel{\mathop{\Big|_{\beta=1,~\tilde{v}=v}}
 _{\mathrm{\tilde{h}=h,~\tilde{q}=q}}}
=-\frac{8 h \pi  q \rho_c  \psi_o  (1-n \alpha +(1+n)
 \alpha  \psi_o )}{(1+n)^2 \alpha ^2 \xi ^3
 \left(1-\frac{2 (1+n) v \alpha }{\xi }+\frac{4
\pi  q^2 \rho_c }{(1+n) \alpha  \xi ^2}\right)}\\\notag
&&-\frac{8 h \pi  q \rho_c  \psi_o  (1-n \alpha +(1+n) \alpha
\psi_o ) \left(v \xi -\frac{4 \pi  q^2 \rho_c }{(1+n)^2 \alpha ^2}
+\alpha  \psi_o ^{1+n}\right)}{(1+n)
\alpha  \xi ^5 \left(1-\frac{2 (1+n) v \alpha }{\xi }
+\frac{4 \pi  q^2 \rho_c }{(1+n) \alpha  \xi ^2}\right)^2}\\
&&-\frac{4 \pi  \rho_c  }{(1+n)^3 \alpha
^3 \xi ^4}\frac{d q}{d \xi}-\frac{4 \pi  q \rho_c }
{(1+n)^3 \alpha ^3 \xi ^4}\frac{d^2 q}{d \xi^2}.
\end{eqnarray}
Furthermore, we have
\begin{eqnarray}\label{36}
\tilde{v}=\frac{1}{\rho_c}\int_0^{\xi}\Big[\bar{\xi}^2
\big\{\rho_{oc}\psi_o^n+n\beta\psi_0^{n+1}P_{rc}\big\}+
\frac{4\pi\rho_c}{\alpha^2 (n+1)^2}\big\{\frac{q}{\bar{\xi}}
\frac{d q}{d \bar{\xi}}-\frac{q^2}{\bar{\xi}^2} \big\} \Big]d \bar{\xi},
\end{eqnarray}
and
\begin{eqnarray}\label{37}
\delta v=\Big(\frac{\partial\hat{\tilde{v}}}
{\partial \beta}\Big) \mathrel{\mathop{\Big|_{\beta=1}}}\delta \beta,~~~~
\delta q=\Bigg(\frac{\frac{\partial\hat{\tilde{v}}}
{\partial \beta}}{\frac{\partial\hat{\tilde{v}}}{\partial q}}\Bigg)
\mathrel{\mathop{\Big|_{\beta=1}}}\delta \beta,
\end{eqnarray}
which can be written as
\begin{eqnarray}\label{38}
\delta v=n \alpha F_1(\xi)\delta\beta,~~~~
\delta q=\frac{n (n+1)^2 \alpha^3 F_1}{4\pi F_2} \delta\beta,
\end{eqnarray}
where
\begin{eqnarray}\label{39}
 F_1=\int_0^{\xi} \bar{\xi}^2\psi_o^{n+1} d \bar{\xi},~~~
 F_2=\int_0^{\xi} \big\{-\frac{2 q}{\bar{\xi}^2}
  +\frac{1}{\bar{\xi}}( \frac{d q}{d \bar{\xi}}
  +\frac{q}{\frac{d q}{d \bar{\xi}}}
  \frac{d^2 q}{d \bar{\xi}^2}) \big\} d \bar{\xi}.
\end{eqnarray}

Now for cracking (overturning) to occur, there
must be a change in sign in $\delta\hat{\tilde{R}}$. More
specifically, it should be positive in the inner regions and
negative at the outer once, i.e.,
$\delta\hat{\tilde{R}}=0$ for some value of
$\xi$ in the interval $[0, \xi_\Sigma]$, implying in turn
\begin{eqnarray}\label{40}
\delta h=-\Gamma\delta\beta,
\end{eqnarray}
where
\begin{eqnarray}\label{41}
\Gamma=\Bigg(
\frac{\frac{n(n+1)^2\alpha^3 F_1}{4\pi F_2}
\frac{\partial\hat{\tilde{R}}}{\partial q}
+n\alpha F_1\frac{\partial\hat{\tilde{R}}}
{\partial \hat{\tilde{v}}}
+\frac{\partial\hat{\tilde{R}}}{\partial \beta} }
{\frac{\partial\hat{\tilde{R}}}{\partial h}}\Bigg)
\mathrel{\mathop{\Big|_{\beta=1,~\tilde{v}=v}}
_{\mathrm{\tilde{h}=h,~\tilde{q}=q}}}.
\end{eqnarray}
Thus, Eq. (\ref{32}) along with above equations implies
\begin{eqnarray}\label{42}\notag
&&\delta\hat{\tilde{R}}=\Bigg\{-
\frac{\Gamma  \psi_o  (1-n \alpha +(1+n) \alpha  \psi_o )
 (v \xi -\frac{4 \pi  q^2 \rho_c }{(1+n)^2 \alpha ^2}
 +\alpha \psi_o ^{1+n})}
{\xi ^3 (1-\frac{2 (1+n) v \alpha }{\xi }+
\frac{4 \pi  q^2 \rho_c }{(1+n) \alpha  \xi ^2})}+\\ \notag &&
n \alpha  F_1\Bigg(\frac{h \psi_o
(1-n \alpha +(1+n) \alpha  \psi_o )}{\xi ^2
(1-\frac{2 (1+n) v \alpha }{\xi }+\frac{4 \pi  q^2 \rho_c
}{(1+n) \alpha  \xi ^2})}+\\ \notag &&
\frac{2 h (1+n) \alpha  \psi_o  (1-n \alpha +(1+n)
\alpha  \psi_o ) (v \xi -\frac{4 \pi  q^2 \rho_c }
{(1+n)^2 \alpha ^2}+\alpha  \psi_o ^{1+n})}{\xi ^4
(1-\frac{2 (1+n) v \alpha }{\xi }+\frac{4 \pi  q^2
\rho_c }{(1+n) \alpha  \xi ^2})^2}\Bigg) +\\ \notag &&
\Bigg(\frac{h \alpha  \psi_o ^{2+n} (1-n \alpha +(1+n)
\alpha  \psi_o )}{\xi ^3 (1-\frac{2 (1+n) v \alpha }{\xi }
+\frac{4 \pi  q^2 \rho_c }{(1+n) \alpha  \xi ^2})}
+\frac{h (1+n) \alpha  \psi_o ^2 (v \xi -\frac{4 \pi
q^2 \rho_c }{(1+n)^2 \alpha ^2}+\alpha  \psi_o ^{1+n})}{\xi
^3 (1-\frac{2 (1+n) v \alpha }{\xi }+\frac{4 \pi
 q^2 \rho_c }{(1+n) \alpha  \xi ^2})}\\ \notag &&+\psi_o ^n
  \frac {d \psi_o}{d \xi} \Bigg)+\frac{n (1+n)^2 \alpha ^3  F_1}{4 \pi  F_2 }\Bigg
(-\frac{8 h \pi  q \rho_c  \psi_o  (1-n \alpha +(1+n) \alpha  \psi_o )}
{(1+n)^2 \alpha ^2 \xi ^3 (1-\frac{2
(1+n) v \alpha }{\xi }+\frac{4 \pi  q^2 \rho_c }{(1+n) \alpha  \xi ^2})}
\\ \notag
&&
-\frac{8 h \pi  q \rho_c  \psi_o  (1-n \alpha +(1+n) \alpha  \psi_o ) (v
\xi -\frac{4 \pi  q^2 \rho_c }{(1+n)^2 \alpha ^2}+\alpha  \psi_o ^{1+n})}
{(1+n) \alpha  \xi ^5 (1-\frac{2 (1+n) v \alpha }{\xi }+\frac{4 \pi
 q^2 \rho_c }{(1+n) \alpha  \xi ^2})^2}
 \\
&&-\frac{4 \pi  \rho_c \frac{d q}{d \xi} }{(1+n)^3
\alpha ^3 \xi ^4}-\frac{4 \pi  q \rho_c
\frac{d^2 q}{d \xi^2} }{(1+n)^3 \alpha ^3 \xi ^4}\Bigg)\Bigg\}\delta \beta.
\end{eqnarray}
It would be more convenient to use variable $x$, defined by
\begin{equation}\label{43}
\xi=\bar{A}x,  ~~~\bar{A}=A r_\Sigma=\xi_\Sigma,
\end{equation}
in terms of which $(\ref{42})$ becomes
\begin{eqnarray}\label{44}\notag
&&\delta\hat{\tilde{R}}=\Bigg\{-
\frac{\Gamma  \psi_o  (1-n \alpha +(1+n) \alpha  \psi_o )
(v (\bar{A}x) -\frac{4 \pi  q^2 \rho_c }
{(1+n)^2 \alpha ^2}+\alpha \psi_o ^{1+n})}
{(\bar{A}x) ^3 (1-\frac{2 (1+n) v \alpha }
{(\bar{A}x) }+\frac{4 \pi  q^2 \rho_c }{(1+n) \alpha  (\bar{A}x) ^2})}+
\\ \notag && n \alpha  F_1 \Bigg(\frac{h \psi_o
(1-n \alpha +(1+n) \alpha  \psi_o )}{(\bar{A}x) ^2
(1-\frac{2 (1+n) v \alpha }{(\bar{A}x) }+\frac{4 \pi  q^2 \rho_c
}{(1+n) \alpha  (\bar{A}x) ^2})}+
\\ \notag &&
\frac{2 h (1+n) \alpha  \psi_o
(1-n \alpha +(1+n) \alpha  \psi_o )
(v (\bar{A}x) -\frac{4 \pi  q^2 \rho_c }{(1+n)^2 \alpha
^2}+\alpha  \psi_o ^{1+n})}{(\bar{A}x) ^4
(1-\frac{2 (1+n) v \alpha }{(\bar{A}x) }+\frac{4 \pi
 q^2 \rho_c }{(1+n) \alpha  (\bar{A}x) ^2})^2}\Bigg) +\\ \notag
&&\Bigg(\frac{h \alpha  \psi_o ^{2+n} (1-n \alpha +(1+n)
\alpha  \psi_o )}{(\bar{A}x) ^3 (1-\frac{2 (1+n) v \alpha }
{(\bar{A}x) }+\frac{4 \pi  q^2
\rho_c }{(1+n) \alpha  (\bar{A}x) ^2})}+\frac{h (1+n) \alpha
\psi_o ^2 (v (\bar{A}x) -\frac{4 \pi  q^2 \rho_c }{(1+n)^2
\alpha ^2}+\alpha  \psi_o ^{1+n})}{(\bar{A}x)
^3 (1-\frac{2 (1+n) v \alpha }{(\bar{A}x) }+\frac{4 \pi
 q^2 \rho_c }{(1+n) \alpha  (\bar{A}x) ^2})}
\\ \notag
&&+\psi_o ^n \frac{1}{\bar{A}}\frac {d \psi_o}{d x}
\Bigg)+\frac{n (1+n)^2 \alpha ^3  F_1}{4 \pi  F_2 }
\Bigg
(-\frac{8 h \pi  q \rho_c  \psi_o  (1-n \alpha +(1+n)
\alpha  \psi_o )}
{(1+n)^2 \alpha ^2 (\bar{A}x) ^3 (1-\frac{2
(1+n) v \alpha }{(\bar{A}x) }+\frac{4 \pi  q^2 \rho_c
}{(1+n) \alpha  (\bar{A}x) ^2})}\\ \notag
&&-\frac{8 h \pi  q \rho_c  \psi_o  (1-n \alpha
+(1+n) \alpha  \psi_o ) (v
(\bar{A}x) -\frac{4 \pi  q^2 \rho_c }{(1+n)^2
\alpha ^2}+\alpha  \psi_o ^{1+n})}
{(1+n) \alpha  (\bar{A}x) ^5 (1-\frac{2 (1+n) v
\alpha }{(\bar{A}x) }+\frac{4 \pi
 q^2 \rho_c }{(1+n) \alpha  (\bar{A}x) ^2})^2}
 \\
&&-\frac{4 \pi  \rho_c \frac{1}{\bar{A}}\frac{d q}{d x} }
{(1+n)^3 \alpha ^3 (\bar{A}x) ^4}-\frac{4 \pi  q \rho_c
\frac{1}{\bar{A}^2}\frac{d^2 q}{d x^2} }{(1+n)^3 \alpha ^3
(\bar{A}x) ^4}\Bigg)\Bigg\}\delta \beta.
\end{eqnarray}
Another perturbation scheme may be used which is
based on the perturbations of energy density and
anisotropy through parameters $n,~h$ and $q$.
Using the same arguments as
in the previous scheme, we have
\begin{eqnarray}\label{45}
P_r=K\rho_o^{1+\frac{1}{n}}=K\rho_{oc}^{1+\frac{1}{n}}\psi_o^{1+n},
\\\label{46}
\rho=\rho_o+nP_r=\rho_{oc}\psi_o^n +
 n k \rho_{oc}^{1+\frac{1}{n}} \psi_o^{1+n},
\\\label{47}
n\longrightarrow \tilde{n}+\delta n,
~~~~~q\longrightarrow \tilde{q}+\delta q,~~~~~
h\longrightarrow \tilde{h}+\delta h.
\end{eqnarray}
Here, we assume that the radial pressure remains
unchanged under perturbation, so
\begin{equation}\label{49}
\tilde{P_r}=P_r=K\rho_{oc}^{1+\frac{1}{n}}\psi_o^{1+n},
\end{equation}
and
\begin{equation}\label{50}
\tilde{\rho}=\rho_{oc}\psi_o^{1+\tilde{n}}+\tilde{n}P_r.
\end{equation}
Thus, hydrostatic equilibrium equation becomes
\begin{eqnarray}\label{51}\notag
\hat{\tilde{R}}&=& \psi_o^n \frac{d \psi_o}{d \xi}
+\frac{\tilde{h}}{\xi^3}((1-n\alpha)\psi_o^{\tilde{n}}
+(\tilde{n}+1)\alpha\psi_o^{n+1})\\
&\times&\Big[\frac{\xi\tilde{v}-\frac{4\pi\rho_c
\tilde{q}^2}{(n+1)^2 \alpha^2}+ \alpha \xi^4
\psi_o^{(n+1)}}{1-2(n+1)\alpha \frac{\tilde{v}}{\xi}+
\frac{4\pi\rho_c \tilde{q}^2}{(n+1) \alpha\xi^2}}\Big]
-\frac{4\pi\rho_c }{(n+1)^3 \alpha^3}
\frac{\tilde{q}}{\xi^4}\frac{d \tilde{q}}{d \xi}.
\end{eqnarray}
Now follows the same scheme used above, we have
\begin{eqnarray}\label{52}\notag
\delta\hat{R}&=&\Big(\frac{\partial\hat{\tilde{R}}}
{\partial \tilde{n}}\Big)
\mathrel{\mathop{\Big|_{\tilde{n}=n,~\tilde{v}=v}}_
{\mathrm{\tilde{h}=h,~\tilde{q}=q}}} \delta\beta
+ \Big(\frac{\partial\hat{\tilde{R}}}{\partial \tilde{v}}\Big)
\mathrel{\mathop{\Big|_{\tilde{n}=n,~\tilde{v}=v}}_
{\mathrm{\tilde{h}=h,~\tilde{q}=q}}} \delta v\\
&&+ \Big(\frac{\partial\hat{\tilde{R}}}
{\partial \tilde{h}}\Big) \mathrel{\mathop{\Big
|_{\tilde{n}=n,~\tilde{v}=v}}_{\mathrm{\tilde{h}=h,
~\tilde{q}=q}}} \delta h+ \Big(\frac{\partial\hat{\tilde{R}}}
{\partial \tilde{q}}\Big) \mathrel{\mathop{\Big
|_{\tilde{n}=n,~\tilde{v}=v}}_{\mathrm{\tilde{h}=h,~\tilde{q}=q}}} \delta q,
\end{eqnarray}
with
\begin{eqnarray}\label{53}
\tilde{v}=\int_0^{\xi}\Big[\bar{\xi}^2\big\{
(1-n\alpha)\psi_o^{\tilde{n}}+\tilde{n}\alpha\psi_0^{n+1}
\big\}
+\frac{4\pi\rho_c}{\alpha^2 (n+1)^2}\big\{\frac{q}
{\bar{\xi}}\frac{d q}{d \bar{\xi}}-\frac{q^2}
{\bar{\xi}^2}\big\} \Big]d \bar{\xi},
\end{eqnarray}
and
\begin{eqnarray}\label{54}
F_3 =\int_0^{\xi} \bar{\xi}^2\big\{(1-n\alpha)
\psi_o^{\tilde{n}}ln \psi_o+ \alpha\psi_0^{n+1}\big\}d \bar{\xi}.
\end{eqnarray}
Again, for the cracking to occur, we must have
$\delta\hat{\tilde{R}}=0$ for some value of
$\xi$ in the interval $[0, \xi_\Sigma]$, implying in turn
\begin{eqnarray}\label{55}
\delta h=-\Gamma\delta n,
\end{eqnarray}
where
\begin{eqnarray}\label{56}
\Gamma=\Bigg(
\frac{\frac{(n+1)^2\alpha^2 F_3}{4\pi\rho_c F_2}
\frac{\partial\hat{\tilde{R}}}{\partial q} +
F_1\frac{\partial\hat{\tilde{R}}}{\partial \hat{\tilde{v}}}
+\frac{\partial\hat{\tilde{R}}}{\partial  \tilde{n}} }
{\frac{\partial\hat{\tilde{R}}}{\partial h}}\Bigg)
\mathrel{\mathop{\Big|_{ \tilde{n}=n,~\tilde{v}=v}}
_{\mathrm{\tilde{h}=h,~\tilde{q}=q}}},
\end{eqnarray}
and
\begin{eqnarray}\label{57}\notag
&&\delta\hat{\tilde{R}}=\Bigg\{-\frac{\Gamma   \left(v \xi
-\frac{4 \pi  q^2 \rho_c }{(1+n)^2 \alpha ^2}+\alpha
\psi_o ^{1+n}\right) \left((1-n \alpha
) \psi_o ^n+(1+n) \alpha  \psi_o ^{1+n}\right)}{\xi ^3
\left(1-\frac{2 (1+n) v \alpha }{\xi }+\frac{4 \pi  q^2
\rho_c }{(1+n) \alpha  \xi ^2}\right)}+\\ \notag &&
\frac{h  \left(v \xi -\frac{4 \pi  q^2 \rho_c }{(1+n)^2
\alpha ^2}+\alpha  \psi_o ^{1+n}\right) \left(\alpha
\psi_o ^{1+n}+(1-n \alpha) \psi_o ^n \text{ln}[\psi_o
]\right)}{\xi ^3 \left(1-\frac{2 (1+n) v \alpha }{\xi }
+\frac{4 \pi  q^2 \rho_c }{(1+n) \alpha  \xi ^2}\right)}
+\\ \notag &&
 \left(\frac{h \left((1-n \alpha ) \psi_o ^n+(1+n) \alpha
 \psi_o ^{1+n}\right)}{\xi ^2 \left(1-\frac{2 (1+n) v \alpha }{\xi }+\frac{4
\pi  q^2 \rho_c }{(1+n) \alpha  \xi ^2}\right)}+\right.\\ \notag &&
\left.\frac{2 h (1+n) \alpha  \left(v \xi -\frac{4 \pi  q^2 \rho_c }
{(1+n)^2 \alpha ^2}+\alpha  \psi_o ^{1+n}\right) \left((1-n \alpha )
 \psi_o ^n+(1+n)
\alpha  \psi_o ^{1+n}\right)}{\xi ^4 \left(1-\frac{2 (1+n) v
\alpha }{\xi }+\frac{4 \pi  q^2 \rho_c }{(1+n) \alpha  \xi ^2}
\right)^2}\right) F_3+\\ \notag &&
\frac{1}{(1+n)^2 \alpha ^2 F_2}4 \pi   \rho_c  F_3 \left(-\frac{8 h
 \pi  q \rho_c  \left((1-n \alpha ) \psi_o ^n+(1+n) \alpha  \psi_o ^{1+n}\right)}{(1+n)^2
\alpha ^2 \xi ^3 \left(1-\frac{2 (1+n) v \alpha }{\xi }+
\frac{4 \pi  q^2 \rho_c }{(1+n) \alpha  \xi ^2}\right)}-\right.\\ \notag &&
\frac{8 h \pi  q \rho_c  \left(v \xi -\frac{4 \pi  q^2
\rho_c }{(1+n)^2 \alpha ^2}+\alpha  \psi_o ^{1+n}\right)
 \left((1-n \alpha ) \psi_o ^n+(1+n) \alpha
 \psi_o ^{1+n}\right)}{(1+n) \alpha  \xi ^5
 \left(1-\frac{2 (1+n) v \alpha }{\xi }+\frac{4 \pi
 q^2 \rho_c }{(1+n) \alpha  \xi ^2}\right)^2}-\\  &&
\left.\frac{4 \pi  \rho_c \frac{d q}{d \xi}}{(1+n)^3
\alpha ^3 \xi ^4}-\frac{4 \pi  q \rho_c
\frac{d^2 q}{d \xi^2}}{(1+n)^3 \alpha ^3 \xi ^4}\right)
\Bigg\}\text{$\delta $n},
\end{eqnarray}
or
\begin{eqnarray}\label{58}\notag
&&\delta\hat{\tilde{R}}=\Bigg\{-\frac{\Gamma
\left(v (\bar{A}x) -\frac{4 \pi  q^2 \rho_c }
{(1+n)^2 \alpha ^2}+\alpha  \psi_o ^{1+n}\right)
 \left((1-n \alpha
) \psi_o ^n+(1+n) \alpha  \psi_o ^{1+n}\right)}
{(\bar{A}x) ^3 \left(1-\frac{2 (1+n) v \alpha }
{(\bar{A}x) }+\frac{4 \pi  q^2 \rho_c }{(1+n)
\alpha  (\bar{A}x) ^2}\right)}+\\ \notag &&
\frac{h  \left(v (\bar{A}x) -\frac{4 \pi  q^2
\rho_c }{(1+n)^2 \alpha ^2}+\alpha  \psi_o ^{1+n}\right)
\left(\alpha  \psi_o ^{1+n}+(1-n \alpha
) \psi_o ^n \text{ln}[\psi_o ]\right)}{(\bar{A}x) ^3
\left(1-\frac{2 (1+n) v \alpha }{(\bar{A}x) }
+\frac{4 \pi  q^2 \rho_c }{(1+n) \alpha  (\bar{A}x) ^2}\right)}+\\ \notag &&
 \left(\frac{h \left((1-n \alpha ) \psi_o ^n+(1+n) \alpha
  \psi_o ^{1+n}\right)}{(\bar{A}x) ^2 \left(1-\frac{2
 (1+n) v \alpha }{(\bar{A}x) }+\frac{4\pi  q^2 \rho_c }
 {(1+n) \alpha  (\bar{A}x) ^2}\right)}+\right.\\ \notag &&
\left.\frac{2 h (1+n) \alpha  \left(v (\bar{A}x) -\frac{4
\pi  q^2 \rho_c }{(1+n)^2 \alpha ^2}+\alpha  \psi_o ^{1+n}\right)
\left((1-n \alpha ) \psi_o ^n+(1+n)
\alpha  \psi_o ^{1+n}\right)}{(\bar{A}x) ^4 \left(1-\frac{2 (1+n)
v \alpha }{(\bar{A}x) }+\frac{4 \pi  q^2 \rho_c }{(1+n) \alpha
(\bar{A}x) ^2}\right)^2}\right) F_3+\\ \notag &&
\frac{1}{(1+n)^2 \alpha ^2 F_2}4 \pi   \rho_c  F_3 \left(-\frac{8
h \pi  q \rho_c  \left((1-n \alpha ) \psi_o ^n+(1+n) \alpha
\psi_o ^{1+n}\right)}{(1+n)^2
\alpha ^2 (\bar{A}x) ^3 \left(1-\frac{2 (1+n) v \alpha }
{(\bar{A}x) }+\frac{4 \pi  q^2 \rho_c }{(1+n) \alpha
(\bar{A}x) ^2}\right)}-\right.\\ \notag &&
\frac{8 h \pi  q \rho_c  \left(v (\bar{A}x) -\frac{4 \pi  q^2 \rho_c }
{(1+n)^2 \alpha ^2}+\alpha  \psi_o ^{1+n}\right) \left((1-n \alpha )
\psi_o ^n+(1+n) \alpha
 \psi_o ^{1+n}\right)}{(1+n) \alpha  (\bar{A}x) ^5
 \left(1-\frac{2 (1+n) v \alpha }{(\bar{A}x) }+\frac{4 \pi  q^2 \rho_c }
 {(1+n) \alpha  (\bar{A}x) ^2}\right)^2}-\\ &&
\left.\frac{4 \pi  \rho_c \frac{1}{\bar{A}}\frac{d q}{d x}}{(1+n)^3
\alpha ^3 (\bar{A}x) ^4}-\frac{4 \pi  q \rho_c \frac{1}{\bar{A}^2}
\frac{d^2 q}{d x^2}}{(1+n)^3 \alpha ^3 (\bar{A}x) ^4}\right)
\Bigg\}\text{$\delta $n}.
\end{eqnarray}

\subsection{Polytropes for case 2}

Here, we consider the ploytropic EoS of the form
\begin{equation}\label{59}
P_r= K\rho^{1+\frac{1}{n}},~~~
\rho=\rho_o \big(1-K \rho_o^{\frac{1}{n}}\big)^{-n},
\end{equation}
where $\rho$ is the total energy density.
Using EoS $(\ref{59})$ along with assumptions $\rho=\rho_c\psi^n$ and of $(\ref{22})$,
the hydrostatic equilibrium equation (\ref{19}) implies
\begin{eqnarray}\label{61} \notag
&&\frac{d \psi}{d \xi}+\frac{h}{\xi^3}(1+\alpha\psi)
\Big[\frac{v(\xi)\xi+\alpha \xi^4 \psi_o^{n+1}
-\frac{4\pi \rho_c q^2}{\alpha^2 (n+1)^2}}{1-2\alpha (n+1)
\frac{v(\xi)}{\xi}+\frac{4\pi\rho_c q^2}{\alpha (n+1)\xi^2}}\Big]
\\&&-\frac{4\pi\rho_c q }{\alpha^3 (n+1)^3\xi^4 }\frac{1}{\psi^n}\frac{d q}{d \xi}=0.
\end{eqnarray}
Differentiating Eq. (\ref{16}) with respect
to $``r"$ and using the assumptions given in Eq. (\ref{22}), we get
\begin{eqnarray}\label{62}
\frac{dv(\xi)}{d\xi}=\xi^2\psi^n+\frac{4\pi\rho_c}
{\alpha^2 (n+1)^2}\Big(\frac{q}{\xi}\frac{d q}
{d \xi}-\frac{q^2}{\xi^2}\Big).
\end{eqnarray}
Now, following the same procedure and carried perturbation out through
$K,~q$ and $h$, we obtain
\begin{eqnarray}\label{63}\notag
&&\delta \hat{\tilde{R}}=\Bigg\{
-\frac{\Gamma  \psi ^n (1-n \alpha  \psi +(1+n) \alpha  \psi )
(v (\bar{A}x) -\frac{4 \pi  q^2 \rho_c }{(1+n)^2 \alpha ^2}+\alpha
\psi ^{1+n})}{(\bar{A}x) ^3 (1-\frac{2 (1+n) v \alpha }
{(\bar{A}x) }+\frac{4 \pi  q^2 \rho_c }{(1+n) \alpha
(\bar{A}x) ^2})}\\ \notag &&+
n \alpha    \Bigg(\frac{h \psi ^n (1-n \alpha  \psi +(1+n)
\alpha  \psi )}{(\bar{A}x) ^2 (1-\frac{2 (1+n) v \alpha }
{(\bar{A}x) }+\frac{4 \pi
 q^2 \rho_c }{(1+n) \alpha  (\bar{A}x) ^2})}+\\
\notag &&\frac{2 h (1+n) \alpha  \psi ^n (1-n \alpha
\psi +(1+n) \alpha  \psi ) (v (\bar{A}x) -\frac{4 \pi  q^2 \rho_c
}{(1+n)^2 \alpha ^2}+\alpha  \psi ^{1+n})}{(\bar{A}x) ^4
(1-\frac{2 (1+n) v \alpha }{(\bar{A}x) }+
\frac{4 \pi  q^2 \rho_c }{(1+n) \alpha  (\bar{A}x) ^2})^2}\Bigg)
F_4+\\ \notag &&
\Bigg(\frac{h \alpha  \psi ^{1+2 n} (1-n \alpha  \psi +(1+n)
\alpha  \psi )}{(\bar{A}x) ^3 (1-\frac{2 (1+n) v \alpha }
{(\bar{A}x) }+\frac{4 \pi  q^2 \rho_c }{(1+n) \alpha
(\bar{A}x) ^2})}+\\ \notag &&\frac{h (1+n) \alpha
\psi ^{1+n} (v (\bar{A}x) -\frac{4 \pi  q^2 \rho_c }
{(1+n)^2 \alpha ^2}+\alpha  \psi
^{1+n})}{(\bar{A}x) ^3 (1-\frac{2 (1+n) v \alpha }
{(\bar{A}x) }+\frac{4 \pi  q^2 \rho_c }{(1+n) \alpha
(\bar{A}x) ^2})}+\psi ^n \frac{1}{\bar{A} }\frac{d \psi}{d x}\Bigg)
+\\ \notag &&
\frac{n (1+n)^2 \alpha ^3   F_4}{4 \pi  F_2}
\Bigg(-\frac{8 h \pi  q \rho_c  \psi ^n (1-n \alpha
\psi +(1+n) \alpha  \psi )}{(1+n)^2 \alpha ^2 (\bar{A}x)
^3 (1-\frac{2 (1+n) v \alpha }{(\bar{A}x) }+\frac{4 \pi
q^2 \rho_c }{(1+n) \alpha  (\bar{A}x) ^2})}\\ \notag &&
-\frac{8 h \pi  q \rho_c  \psi ^n (1-n \alpha  \psi
+(1+n) \alpha  \psi ) (v (\bar{A}x) -\frac{4 \pi  q^2
\rho_c }{(1+n)^2 \alpha ^2}+\alpha  \psi ^{1+n})}{(1+n)
\alpha  (\bar{A}x) ^5 (1-\frac{2 (1+n)
v \alpha }{(\bar{A}x) }+\frac{4 \pi  q^2 \rho_c }{(1+n)
\alpha  (\bar{A}x) ^2})^2} \\ &&-\frac{4 \pi  \rho_c
\frac{1}{\bar{A} }\frac{d q}{d x}}{(1+n)^3 \alpha ^3
(\bar{A}x) ^4}-\frac{4 \pi  q \rho_c \frac{1}{\bar{A}^2}
\frac{d^2 q }{d x^2} }{(1+n)^3
\alpha ^3 (\bar{A}x) ^4}\Bigg)\Bigg\}\delta \beta,
\end{eqnarray}
and if perturbation is carried out through
$n,~q$ and $h$, then we obtain
\begin{eqnarray}\label{64}\notag
&&\delta \hat{\tilde{R}}=\Bigg\{
-\frac{\Gamma   (v (\bar{A}x) -\frac{4 \pi  q^2 \rho_c }
{(1+n)^2 \alpha ^2}+\alpha  \psi ^{1+n}) (\psi ^n+(2+n)
\alpha  \psi ^{1+n})}{(\bar{A}x) ^3 (1-\frac{2 (1+n) v
\alpha }{(\bar{A}x) }+\frac{4 \pi  q^2 \rho_c }{(1+n)
\alpha  (\bar{A}x) ^2})}+\\ \notag &&
\frac{h  (v (\bar{A}x) -\frac{4 \pi  q^2 \rho_c }{(1+n)^2
\alpha ^2}+\alpha  \psi ^{1+n}) (\alpha  \psi ^{1+n}+\psi ^n
\text{ln}[\psi
])}{(\bar{A}x) ^3 (1-\frac{2 (1+n) v \alpha }{(\bar{A}x) }+
\frac{4 \pi  q^2 \rho_c }{(1+n) \alpha  (\bar{A}x) ^2})}+
\Bigg(\frac{h (\psi ^n+(2+n) \alpha  \psi ^{1+n})}{(\bar{A}x) ^2
(1-\frac{2 (1+n) v \alpha }{(\bar{A}x) }+\frac{4 \pi  q^2 \rho_c
}{(1+n) \alpha  (\bar{A}x) ^2})}\\ \notag &&+\frac{2 h (1+n) \alpha
(v (\bar{A}x) -\frac{4 \pi  q^2 \rho_c }{(1+n)^2 \alpha ^2}+\alpha
\psi ^{1+n}) (\psi^n+(2+n) \alpha  \psi ^{1+n})}{(\bar{A}x) ^4
(1-\frac{2 (1+n) v \alpha }{(\bar{A}x) }+\frac{4 \pi  q^2 \rho_c }
{(1+n) \alpha  (\bar{A}x) ^2})^2}\Bigg)
F_5\\ \notag &&+\frac{4 \pi \rho_c  F_5}{(1+n)^2 \alpha ^2 F_2}
\Bigg(-\frac{8 h \pi  q \rho_c  (\psi ^n+(2+n) \alpha  \psi ^{1+n})}{(1+n)^2
\alpha ^2 (\bar{A}x) ^3 (1-\frac{2 (1+n) v \alpha }{(\bar{A}x) }
+\frac{4 \pi  q^2 \rho_c }{(1+n) \alpha  (\bar{A}x) ^2})}-\\ \notag &&
\frac{8 h \pi  q \rho_c  (v (\bar{A}x) -\frac{4 \pi  q^2 \rho_c }
{(1+n)^2 \alpha ^2}+\alpha  \psi ^{1+n}) (\psi ^n+(2+n) \alpha
\psi^{1+n})}{(1+n) \alpha  (\bar{A}x) ^5 (1-\frac{2 (1+n) v
\alpha }{(\bar{A}x) }+\frac{4 \pi  q^2 \rho_c }{(1+n) \alpha
(\bar{A}x) ^2})^2}-\\ \notag &&\frac{4 \pi  \rho_c
\frac{1}{\bar{A} } \frac{d q}{d  x } }{(1+n)^3 \alpha ^3
(\bar{A}x) ^4}-\frac{4 \pi  q \rho_c \frac{1}{\bar{A}^2}
\frac{d^2 q }{d x^2} }{(1+n)^3 \alpha ^3 (\bar{A}x) ^4}\Bigg)
\Bigg\}\text{$\delta $n},
\end{eqnarray}
where
\begin{eqnarray}\label{65}
 F_4=\int_0^{\xi} \bar{\xi}^2\psi^{n+1} d \bar{\xi},~~~
 F_5 =\int_0^{\xi} \bar{\xi}^2\big\{(\psi^{\tilde{n}}ln
 \psi_o+ \alpha(\beta-1)\psi^{n+1}\big\}d \bar{\xi}.
\end{eqnarray}
%-------------------------------------------------------------------
\begin{figure} \label{fig1}
\centering
\includegraphics[width=80mm]{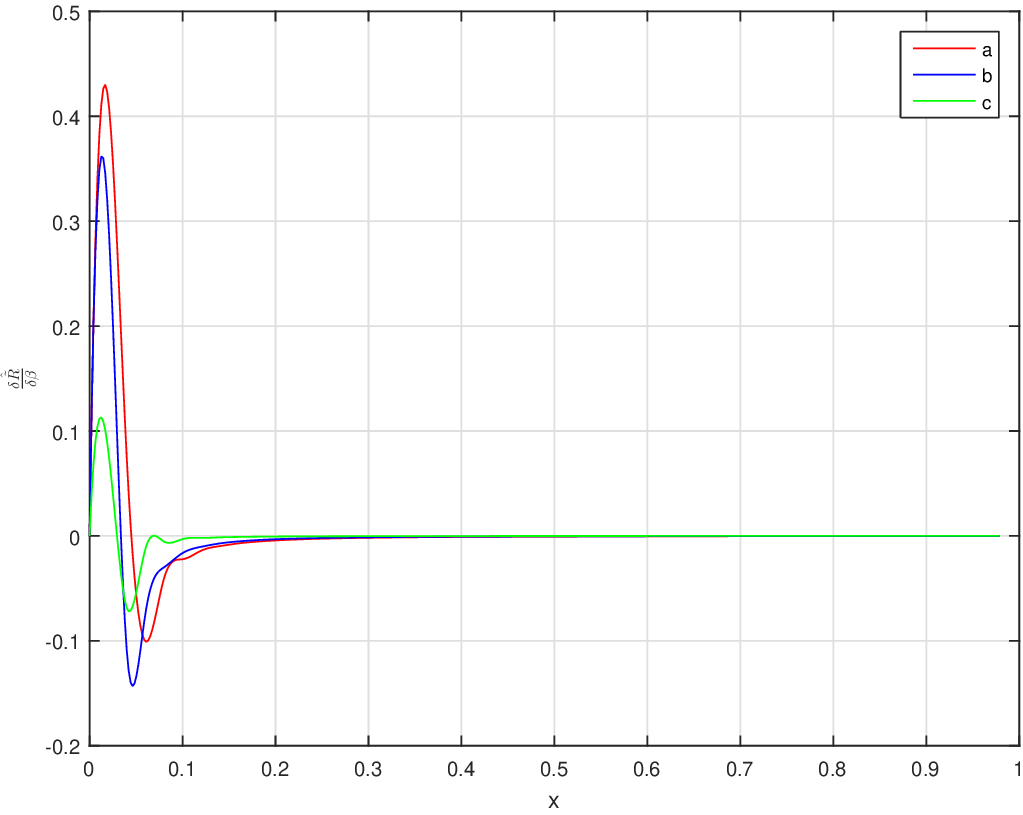}
\caption{Case $1$: Perturbation through $K,~q$ and $h$. $\frac{\delta \hat{\tilde{R}}}{\delta \beta}$ as a function of $x$ for $n=1,~h=1.5,~\Gamma=1.6,~\alpha=0.83$,
curve $a$: Q=0.2 $M_\odot$,
curve $b$: Q=0.4 $M_\odot$,
curve $c$: Q=0.64 $M_\odot$}.
\end{figure}
%-------------------------------------------------------------------
\begin{figure}\label{fig2}
\centering
\includegraphics[width=80mm]{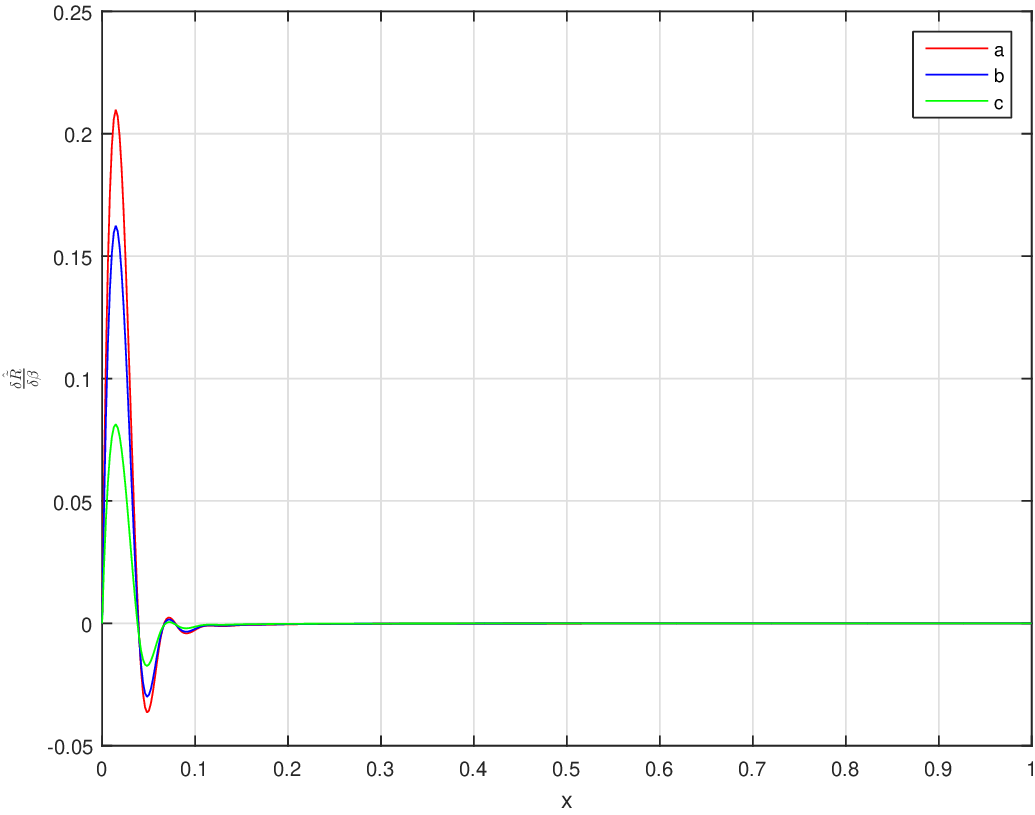}
\caption{Case $1$: Perturbation through $K,~q$ and $h$. $\frac{\delta \hat{\tilde{R}}}{\delta \beta}$ as a function of $x$ for $n=1,~h=1.5,~\Gamma=1.6,~\alpha=0.85$,
curve $a$: Q=0.2 $M_\odot$,
curve $b$: Q=0.4 $M_\odot$,
curve $c$: Q=0.64 $M_\odot$}.
\end{figure}

%-------------------------------------------------------------------
\begin{figure}\label{fig3}
\centering
\includegraphics[width=80mm]{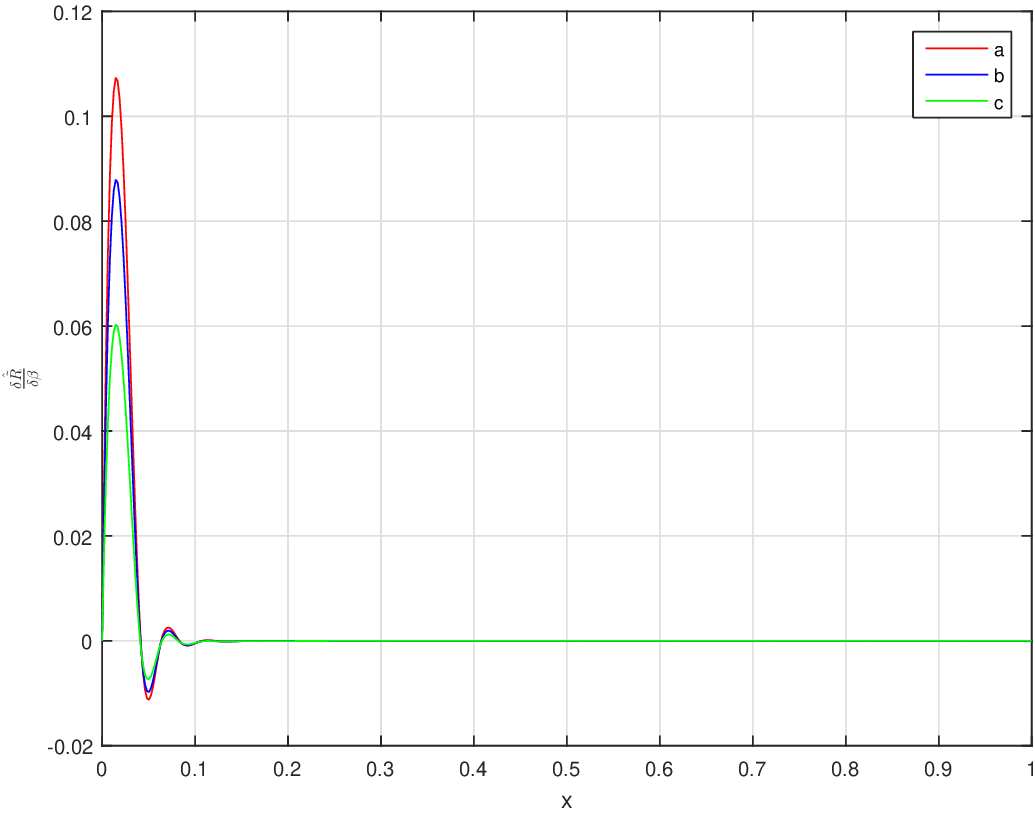}
\caption{Case $1$: Perturbation through $K,~q$ and $h$. $\frac{\delta \hat{\tilde{R}}}{\delta \beta}$ as a function of $x$ for $n=1,~h=1.5,~\Gamma=1.6,~\alpha=0.87$,
curve $a$: Q=0.2 $M_\odot$,
curve $b$: Q=0.4 $M_\odot$,
curve $c$: Q=0.64 $M_\odot$}.
\end{figure}

%-------------------------------------------------------------------
\begin{figure}\label{fig4}
\centering
\includegraphics[width=80mm]{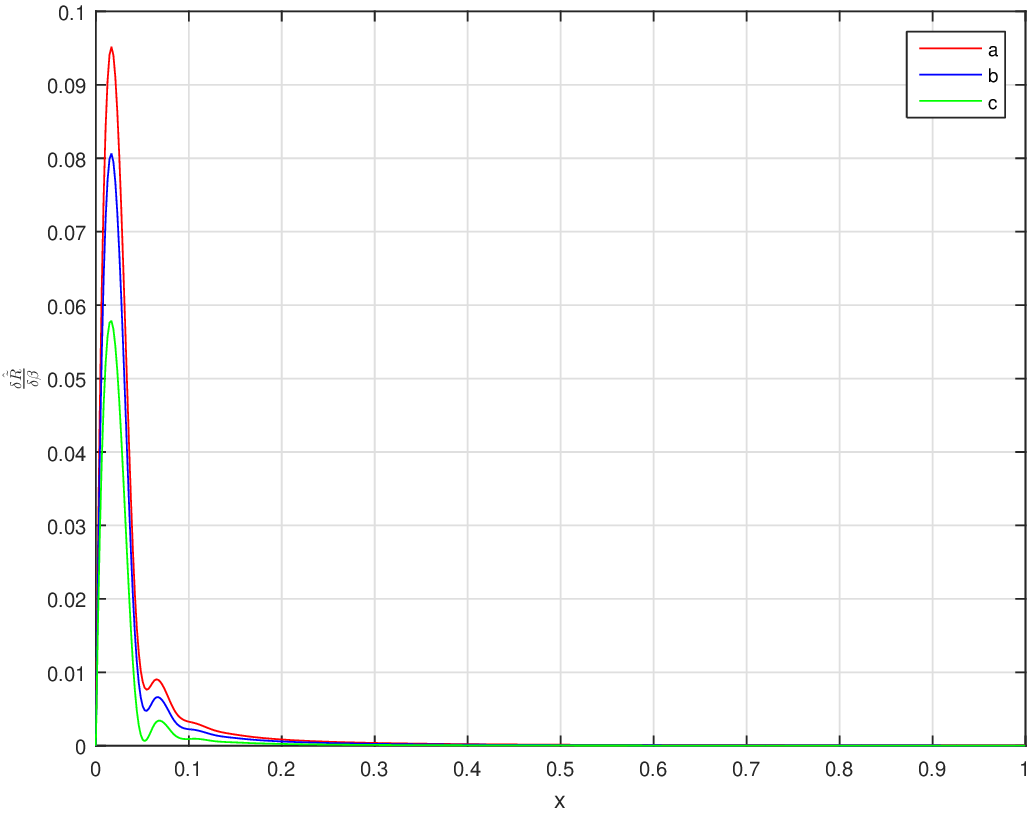}
\caption{Case $1$: Perturbation through $K,~q$ and $h$. $\frac{\delta \hat{\tilde{R}}}{\delta \beta}$ as a function of $x$ for $n=1,~h=1.5,~\Gamma=1.6,~\alpha=0.9$,
curve $a$: Q=0.2 $M_\odot$,
curve $b$: Q=0.4 $M_\odot$,
curve $c$: Q=0.64 $M_\odot$}.
\end{figure}

%-------------------------------------------------------------------
\begin{figure}\label{fig5}
\centering
\includegraphics[width=80mm]{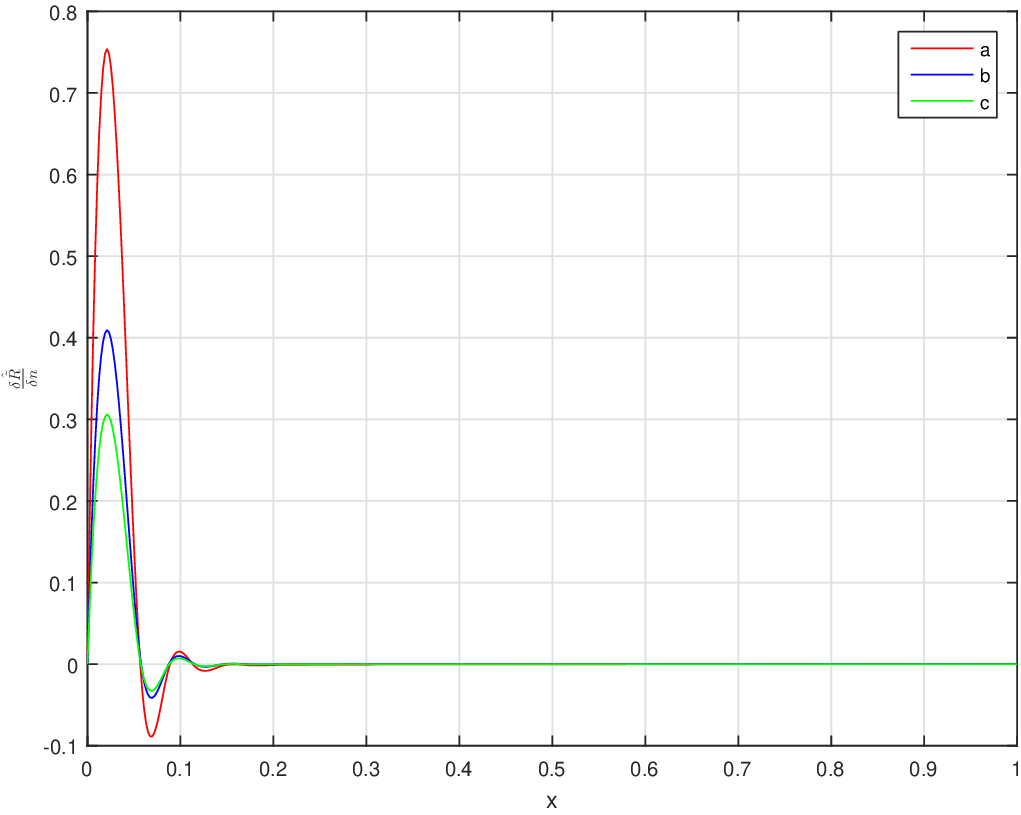}
\caption{Case $1$: Perturbation through $n,~q$ and $h$. $\frac{\delta \hat{\tilde{R}}}{\delta n}$ as a function of $x$ for $n=1,~h=1,~\Gamma=1.2,~\alpha=0.87$,
curve $a$: Q=0.2 $M_\odot$,
curve $b$: Q=0.4 $M_\odot$,
curve $c$: Q=0.64 $M_\odot$}.
\end{figure}

%-------------------------------------------------------------------
\begin{figure}\label{fig6}
\centering
\includegraphics[width=80mm]{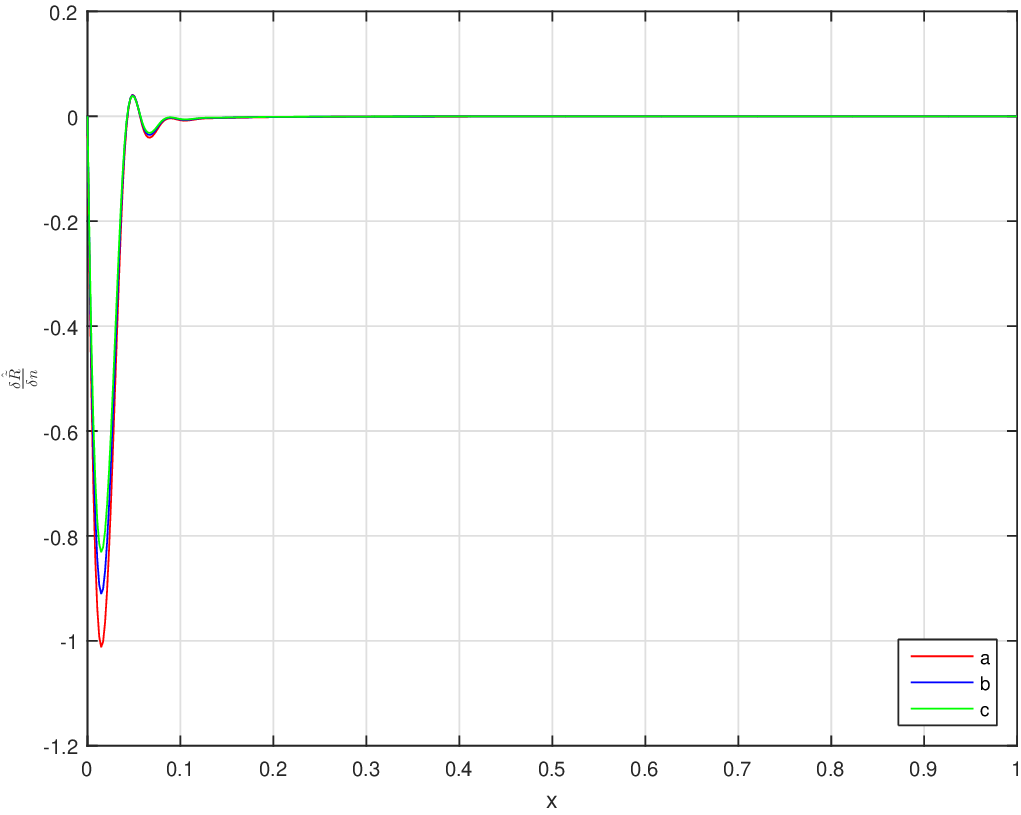}
\caption{Case $1$: Perturbation through $n,~q$ and $h$. $\frac{\delta \hat{\tilde{R}}}{\delta n}$ as a function of $x$ for $n=1.5,~h=1.3,~\Gamma=0.6,~\alpha=0.69$,
curve $a$: Q=0.2 $M_\odot$,
curve $b$: Q=0.4 $M_\odot$,
curve $c$: Q=0.64 $M_\odot$}.
\end{figure}

%-------------------------------------------------------------------
\begin{figure}\label{fig7}
\centering
\includegraphics[width=80mm]{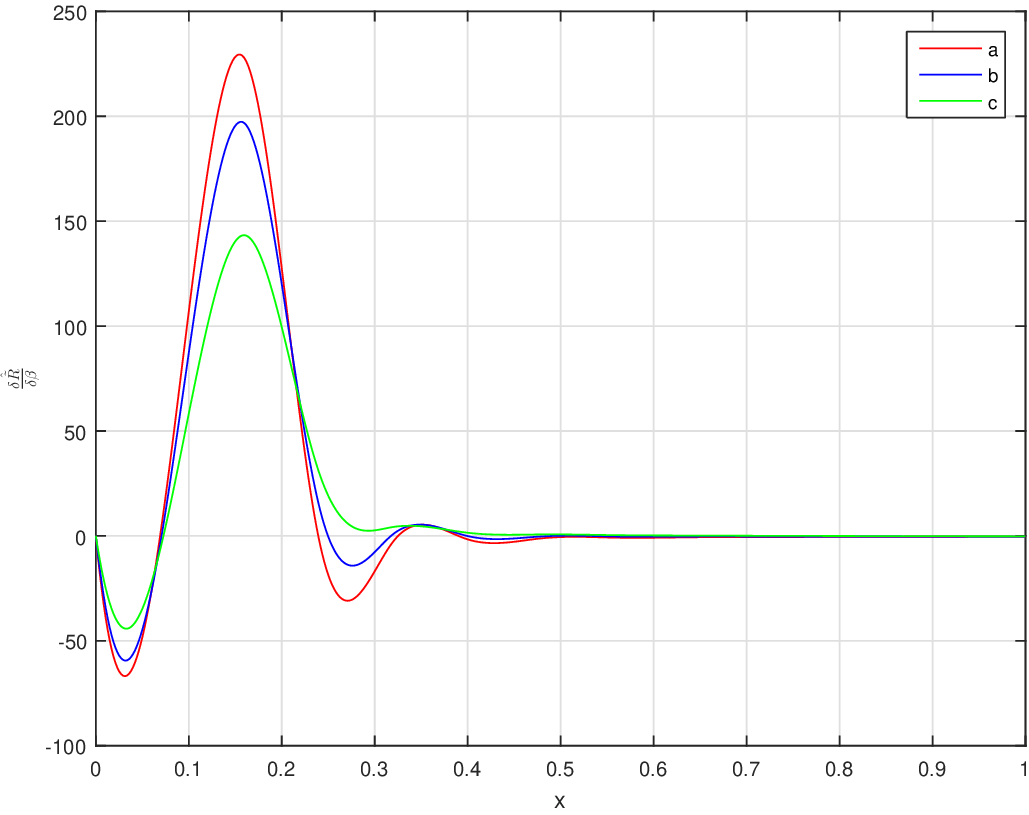}
\caption{Case $2$: Perturbation through $K,~q$ and $h$. $\frac{\delta \hat{\tilde{R}}}{\delta \beta}$ as a function of $x$ for $n=0.5,~h=0.5,~\Gamma=0.6,~\alpha=0.69$,
curve $a$: Q=0.2 $M_\odot$,
curve $b$: Q=0.4 $M_\odot$,
curve $c$: Q=0.64 $M_\odot$}.
\end{figure}

%-------------------------------------------------------------------
\begin{figure}\label{fig8}
\centering
\includegraphics[width=80mm]{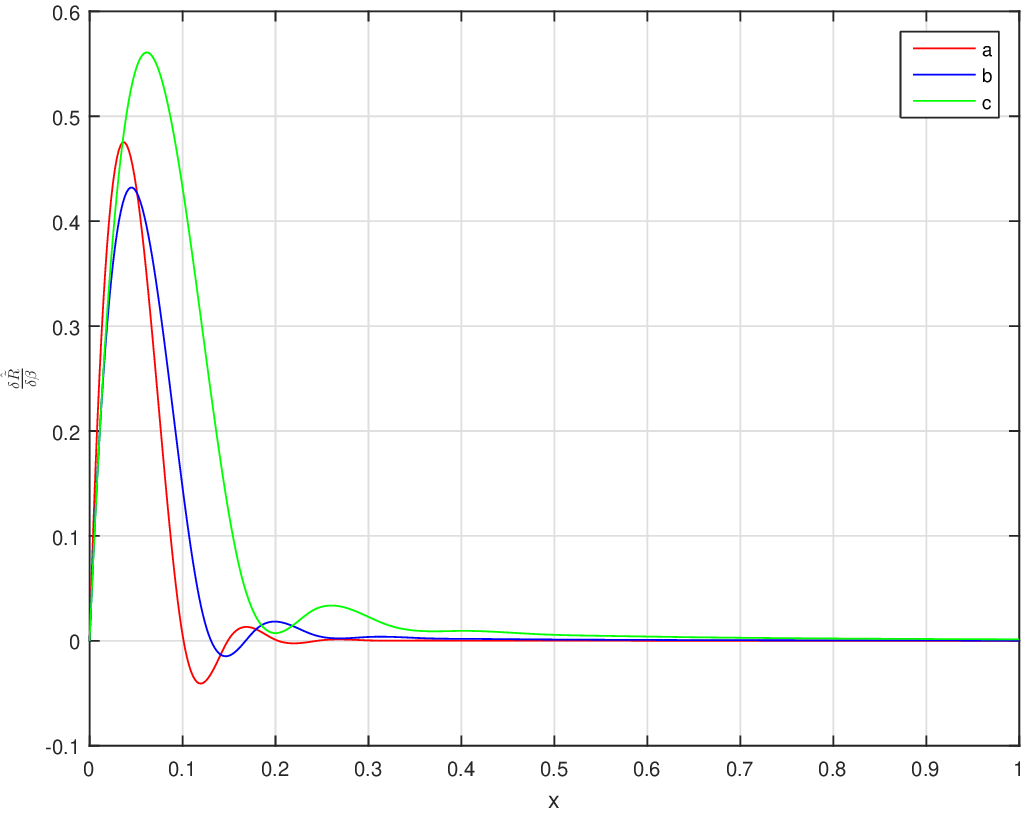}
\caption{Case $2$: Perturbation through $K,~q$ and $h$. $\frac{\delta \hat{\tilde{R}}}{\delta \beta}$ as a function of $x$ for $n=0.5,~h=0.5,~\Gamma=0.6,~\alpha=0.90$,
curve $a$: Q=0.2 $M_\odot$,
curve $b$: Q=0.4 $M_\odot$,
curve $c$: Q=0.64 $M_\odot$}.
\end{figure}

%-------------------------------------------------------------------
\begin{figure}\label{fig9}
\centering
\includegraphics[width=80mm]{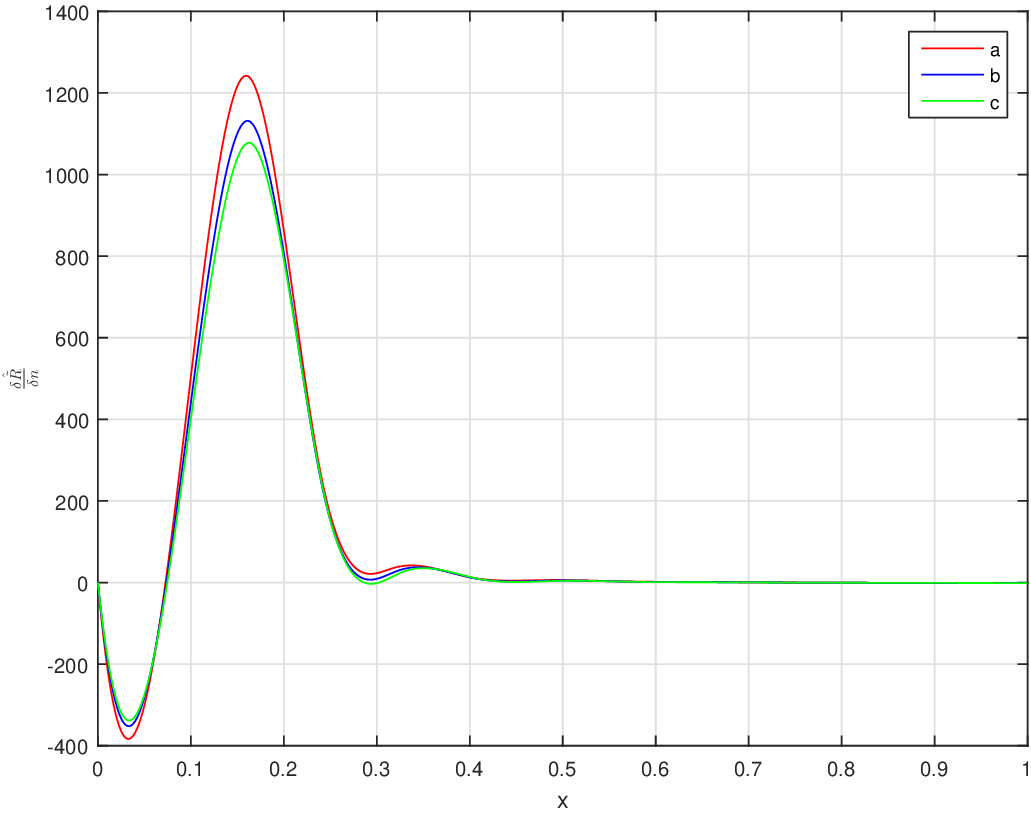}
\caption{Case $2$: Perturbation through $n,~q$ and $h$. $\frac{\delta \hat{\tilde{R}}}{\delta n}$ as a function of $x$ for $n=0.5,~h=0.5,~\Gamma=0.6,~\alpha=0.83$,
curve $a$: Q=0.2 $M_\odot$,
curve $b$: Q=0.4 $M_\odot$,
curve $c$: Q=0.64 $M_\odot$}.
\end{figure}
%-------------------------------------------------------------------
\begin{figure}\label{fig10}
\centering
\includegraphics[width=80mm]{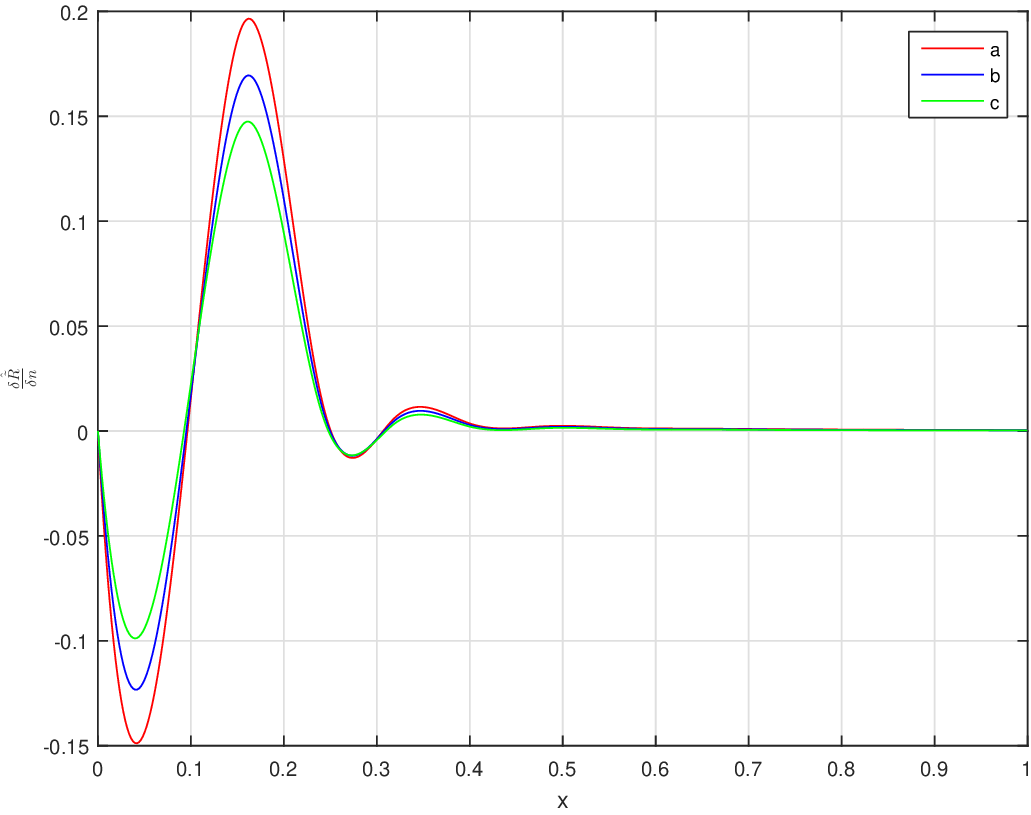}
\caption{Case $2$: Perturbation through $n,~q$ and $h$. $\frac{\delta \hat{\tilde{R}}}{\delta n}$ as a function of $x$ for $n=0.5,~h=0.5,~\Gamma=0.6,~\alpha=0.90$,
curve $a$: Q=0.2 $M_\odot$,
curve $b$: Q=0.4 $M_\odot$,
curve $c$: Q=0.64 $M_\odot$}.
\end{figure}
Now we shall apply the formulism developed in the previous
section to investigate the effects of parameters under perturbation on
the stability of charged anisotropic polytropes defined by
EoS $(\ref{18})$. We shall apply perturbation schemes for
both types of polytropes. For this purpose, fourth-order
Runge-Kutta method has been used for the integration of
Eqs. $(\ref{23}),~(\ref{24})$ and $(\ref{61}),~(\ref{62})$
for any triplet of parameters $n,~\alpha, q$ and $n,~h,~q$
which ensure the exitance of boundary surfaces. Such values
have been suggested in \cite{19,30}. Also, the integrals
in Eqs. $(\ref{39}),~(\ref{54})$ and $(\ref{65})$ were
integrated numerically by using trapezoidal rule.
Next, we have calculated $\psi_o$ and $\psi$ which are used to evaluate
Eqs. $(\ref{44})$ and $(\ref{58})$ for case \textbf{1} and
$(\ref{63})$ and $(\ref{64})$ for case \textbf{2}.

\section{Discussion and conclusion}

Figures \textbf{1-4} summarize the main results for polytropes
of case \textbf{1}, when perturbation is carried out through
parameters $K,~q$ and $h$, whereas figs. \textbf{5} and \textbf{6}
describes the behavior of polytropes when system is perturbed through
parameters $n,~q$ and $h$.
In fig.\textbf{1} the cracking is appeared for $n=1,~h=1.5,~\Gamma=1.6,~\alpha=0.83$
and charge $Q=0.2 M_\odot$, $Q=0.4 M_\odot$ and $ Q=0.64 M_\odot$.
We note that overturning appears in the outer regions of the model.
Figures \textbf{2} and \textbf{3} describes the same behavior for
$\alpha=0.85,~0.9$, respectively. In both cases overturning
is weaker and cracking is stronger. From figs. \textbf{1-4},
we finds that large value of charge $q$ leads towards
weaker overturning and system become gradually stable.
Figure \textbf{4} shows that for $\alpha=0.9$, system become stable.
In fig. \textbf{5} strong cracking and weak overturning
is noted for $n=1,~h=1.0,~\Gamma=1.2,~\alpha=0.87$
and for values of charge $Q=0.2 M_\odot$,
$Q=0.4 M_\odot$ and $ Q=0.64 M_\odot$, whereas
fig. \textbf{6} represent weak cracking and strong overturning
for $n=1.5,~h=1.3,~\Gamma=0.6,~\alpha=0.69$
and for values of charge $Q=0.2 M_\odot$,
$Q=0.4 M_\odot$, $ Q=0.64 M_\odot$, which means system
remain unstable in this case.

Figures \textbf{7} and \textbf{8} sums the main results for polytropes
of case \textbf{2}, when perturbation is carried out through
parameters $K,~q$ and $h$, while figs. \textbf{9} and \textbf{10}
shows cracking when system is perturbed through
parameters $n,~q$ and $h$.
Figures \textbf{7} ($\alpha=0.69$) and  \textbf{8} ($\alpha=0.90$) describe weak overturning near the center
and strong cracking in the outer regions for
$n=0.5,~h=0.5,~\Gamma=0.6$, $ Q=0.2 M_\odot$,
$ Q=0.4 M_\odot$ and $ Q=0.64 M_\odot$ shown
by curves a, b and c respectively. It is noted that
system become stable (see fig.\textbf{8} curve c)
when charge increased from $Q=0.2 M_\odot$ to $Q=0.64 M_\odot$.
Figures \textbf{9} and \textbf{10} describe a similar behavior
when perturbation is carried out through $n,~q$ and $h$
corresponding to some fixed values.

Now, we provide a comparison
between results presented in this work with those
presented in \cite{34}.
The graphical analysis shows that the results obtained
in this work by perturbing charge parameter
differ with the one presented in \cite{34}.
If we compare the results of case \textbf{1}, it is found that
when perturbation is carried out by $K,~h$ and $q$ instead
of $K,~h$, we note deep and strong cracking for small values of
anisotropy factor $h$ and large values of alpha $\alpha$
in the presence of charge (see Figures \textbf{1-3}).
Further, our results (see fig.\textbf{4}) depicts stable configurations
for increasing value of $\alpha$ in the presence of charge.
Also, in case \textbf{1}, it is found that both cracking and overturning
appears when investigation is carried through perturbing $n,~q,~h$ instead of $n,~h$
shown in figs. \textbf{5} and \textbf{6}, whereas the results presented
in \cite{34} shows stable configurations.
A similar results have been obtained for case \textbf{2}, when
perturbation is carried through $K,~q,~h$.
It shows that perturbing the charge parameter
has a significant role on the cracking (or overturning)
of polytropes.

We have developed the general procedure to investigate 
overturning and/or cracking of anisotropic polytropes
through perturbation on anisotropy, energy density
and charge. For both cases, our results have
same behavior. As value of $\alpha$ increases gradually, in general
the system shows strong and deep cracking and weak overturning
for both cases. On the other hand, when charge
increased sufficiently both configurations show stable behavior
under similar conditions.
It should be noted that occurrence of cracking
has immediate effect on stellar structure,
gravitational collapse and evolution of compact objects
but the time scale chosen is much smaller
then the hydrostatic time scale \cite{33}.
Moreover, the discussion of cracking and/or overturning
is the study of snapshot of system just after when it leaves the equilibrium state.
Also with this study one may predict about the system stability by
analyzing the amplitude of cracking (or overturning) which
may lead towards gravitational collapse or expansion of compact objects.
The existence of charge plays an important role in the study
of polytropes. As the amount of charge increase to sufficient extent,
it significantly affect the existence of cracking (or overturning) phenomenon.
In the concluding remarks, we can say that in some regions
the existence of charge may shifts the system from unstable to stable
regions even after perturbation. It is worthwhile to mentioned here that
all our results reduced to \cite{33} for anisotropic spherical polytropes
in the absence of charge.

}

\end{document}